\documentclass[]{spie}  

\usepackage{amsmath,amsfonts,amssymb}
\usepackage{graphicx}
\usepackage[colorlinks=true, allcolors=blue]{hyperref}
\usepackage{glossaries}  
\setacronymstyle{long-short}
\newacronym{PSF}{PSF}{point spread function}
\newacronym{IWA}{IWA}{inner working angle}
\newacronym{HWO}{\textit{HWO}}{\textit{Habitable Worlds Observatory}}
\newacronym{SED}{SED}{spectral energy distribution}
\newacronym{MCMC}{MCMC}{Markov chain Monte Carlo}
\newacronym{SNR}{SNR}{signal-to-noise ratio}

\title{Unveiling habitable planets: Toy coronagraph tackles the exozodiacal dust challenge}

\author[a]{Yu-Chia Lin}
\affil[a]{University of Arizona, Tucson, AZ, USA}

\authorinfo{Further author information: (Send correspondence to Y.L.)\\Y.L.: E-mail: yuchialin@arizona.edu}

\begin{document} 
\maketitle

\begin{abstract}
Directly imaging Earth-like exoplanets within habitable zones is challenging because faint signals can be obscured by exozodiacal dust, analogous to our solar system's zodiacal dust. This dust scatters starlight, creating a bright background noise. This paper introduces Toy Coronagraph, a Python package designed to quantify the impact of this dust on exoplanet detection. It takes circularly symmetric disk \glstarget{PSF} images \& \glspl{PSF}, and exoplanet orbital parameters as input, generating key metrics like contrast curves, signal-to-noise ratios, and dynamic visualizations of exoplanet motion under the dust background. The package also provides tools for generating vortex coronagraph \glspl{PSF} and includes example disk images. Toy Coronagraph empowers researchers to understand exozodiacal dust, develop mitigation strategies, and optimize future telescope designs and mission time, ultimately advancing the search for potentially habitable worlds. Future work will focus on handling non-circularly symmetric inputs, incorporating realistic noise models, and estimating exoplanet yield rates for future space telescope missions.
\end{abstract}


\section{Introduction}
The quest to directly image and characterize Earth-like exoplanets residing in the habitable zones of nearby stars stands as an ambitious objective in contemporary astronomy \cite{2021pdaa.book.....N}. Next-generation space telescopes, such as \glstarget{HWO} the \gls{HWO} \cite{2024arXiv240519418S}, aspire to achieve the extraordinary contrast ratios (on the order of $10^{-10}$) necessary to discern these faint planetary signals from the glare of their host stars \cite{2020arXiv200106683G, 2019arXiv191206219T}. However, this ambitious endeavor faces a significant hurdle from the exozodiacal dust – analogous to the zodiacal dust in our solar system – which can scatter starlight and create a bright background noise that obscures the faint light from exoplanets. Accurately modeling and understanding the impact of exozodiacal dust is, therefore, crucial for the success of future exoplanet direct imaging missions.

To address this critical challenge, I have developed Toy Coronagraph, a versatile Python package designed to simulate and analyze the impact of exozodiacal dust on exoplanet detection capabilities. This open-source tool empowers researchers to explore a variety of scenarios by taking circularly symmetric dust disk images (representing the starlight scattered by the exozodiacal dust) and \glspl{PSF} of the observing system, and exoplanet orbital parameters as inputs. By propagating the disk image through the observing system, Toy Coronagraph generates key metrics essential for evaluating high-contrast imaging systems' performance in the presence of exozodiacal dust. These metrics include contrast curves, signal-to-noise ratios for exoplanet detection, and dynamic visualizations of exoplanet motion in the background of scattered starlight by the exozodiacal dust. Moreover, the package can generate example \glspl{PSF} (ideal vortex coronagraphs) and provide example disk images, offering users a comprehensive platform for investigating the complex interplay between exozodiacal dust and high-contrast imaging techniques. The development of Toy Coronagraph represents a significant step towards achieving a deeper understanding of exozodiacal dust impact, informing the development of effective mitigation strategies, and optimizing the design of future telescopes and the allocation of valuable mission time, ultimately propelling us closer to the discovery of potentially habitable worlds beyond our solar system.

\section{Concepts and Methods}
Accurately modeling the interaction of starlight scattered by exozodiacal dust with a coronagraph's \gls{PSF} is paramount for robust predictions of exoplanet detectability in high-contrast imaging. While existing exoplanet yield estimation software packages often consider the impact of exozodiacal dust, they rely on simplifications that may not fully capture the complexities of this interaction \cite{2019SPIE11117E..0KD}. To address this limitation, the author introduces Toy Coronagraph, a Python package designed to provide a more accurate approach to simulating the impact of exozodiacal dust on exoplanet direct imaging observations.

Toy Coronagraph achieves this by performing a point-by-point equivalent convolution of the input dust disk image with the coronagraph's \glspl{PSF}. Importantly, Toy Coronagraph allows users to explicitly investigate the impact of dust located within \glstarget{IWA} the \gls{IWA}, a region often excluded in simplified simulations. To ensure computational efficiency, Toy Coronagraph currently requires circular symmetry for both the input dust image and the coronagraph's structure which results in circular symmetric \glspl{PSF}.

\subsection{Point-by-point Equivalent Convolution}
Exploiting the circular symmetry of both the \glspl{PSF} and the input disk images, I implement an efficient point-by-point equivalent convolution through rotation. This approach leverages the fact that rotating a circularly symmetric image about its center yields an identical image. However, it is important to account for the unequal weighting of image pixels at different radial distances ($r$) from the center of rotation during this process. To address this, I calculate a pre-rotated image along the positive x-axis using the following equation (excluding the central pixel at $r=0$):

\begin{equation}
\label{eq:cir_preimg}
\text{Pre-Image}(0^{\circ}) = \sum_{r} \frac{2\pi r B(r) \text{PSF}(r)}{N_{\text{rot}}}
\end{equation}

where the pre-rotated image at the initial rotation angle of $0^{\circ}$ is represented by $\text{Pre-Image}(0^{\circ})$. The equation sums the product of the disk image pixel value $B(r)$ and the corresponding \gls{PSF},  $\text{PSF}(r)$, along the positive x-axis, where $r$ denotes the radial distance from the center of the image where the host star is located. The factor $2\pi r$ accounts for the circumference of the circle at radius $r$, ensuring proper weighting of pixels at different radii. This summation is then normalized by the total number of rotation steps, $N_{\text{rot}}$.

Subsequently, the final convolved image is obtained by summing the pre-rotated images at different rotation angles:

\begin{equation}
\label{eq:cir_img}
\text{Image} = \sum_{i=1}^{N_{\text{rot}}} \text{Pre-Image}\left(\frac{i}{N_{\text{rot}}} \times 360^{\circ}\right)
\end{equation}

Here, the summation iterates through the rotation angles, accumulating the contributions from the pre-rotated images at each angle, effectively simulating the convolution of the disk image and the \glspl{PSF}. This process produces the final image representing the exozodiacal dust observed scene through the propagation of a coronagraph.

\subsection{Core Throughput Region}
This package introduces the concept of the "Core Throughput Region" to quantify the spatial extent of a point light source, such as a star or planet, after it passes through a coronagraph. I define the core throughput region as the area encompassing pixels around the brightest pixel where the intensity exceeds half of its peak value. This definition provides a practical and consistent way to characterize the concentrated light from the source.

The core throughput region plays a crucial role in two key aspects of my analysis:
\begin{itemize}
\item \textbf{Characterizing Coronagraph Performance:} I use the core throughput region to calculate the core throughput function of a coronagraph. This function quantifies the fraction of light from a point source that falls within the defined region after being processed by the coronagraph. This metric helps assess the coronagraph's ability to suppress starlight and enhance the visibility of faint exoplanets.

\item \textbf{Estimating the Planet to Dust Brightness Ratio:} The core throughput region enables me to determine the Brightness Ratio between an exoplanet and the exozodiacal dust. By comparing the flux within the exoplanet's core throughput region to the flux from the dust within the same region, I can estimate the relative brightness of the two signals. This ratio is crucial for understanding the detectability of the exoplanet against the dust background, as it directly relates to the \glstarget{SNR} achievable \gls{SNR} in a real observation.
\end{itemize}

\subsection{Features and Functionalities}
Toy Coronagraph provides a range of features designed to facilitate the study of exozodiacal dust:

\begin{itemize}
\item \textbf{Flexible Input Parameter Adjustment:} Users can easily customize input and simulation parameters such as the input FITS file, number of cores for parallel processing, rotation steps for circular symmetry, wavelength, aperture size, Lyot mask size, and \gls{PSF} range. This is achieved through a Python parameter file (toycoronagraph\_para.py), offering a high degree of control over the simulation setup. 

\item \textbf{Realistic Dust Disk Image:} The package includes an example dust disk image, generated for Epsilon Eridani using MCFOST \cite{2006A&A...459..797P} \glstarget{MCMC} with \gls{MCMC} \cite{2013PASP..125..306F} fitting \glstarget{SED} of \gls{SED} observations \cite{2023PASP..135l5001A}. Users can also provide their custom dust images for running simulations.

\item \textbf{Example \gls{PSF} Generation:} Leveraging the HCIPy package \cite{por2018hcipy}, Toy Coronagraph enables users to generate \glspl{PSF} for vortex coronagraphs with different charge numbers, facilitating the exploration of various coronagraph designs. For vortex coronagraph with charge numbers 2, 4, 6, 8, 10, or 12, the corresponding \glspl{PSF} can be found at the release of Toy Coronagraph on GitHub \footnote{\url{https://github.com/dreamjade/Toy_Coronagraph/releases/tag/1.6.1}}.

\item \textbf{Comprehensive Analysis Tools:} The package allows for generating contrast curves, calculating signal-to-noise ratios with and without inner dust, and creating videos showcasing exoplanet motion against the dust background.

\item \textbf{User-Friendly Interface and Comprehensive Documentation:} Toy Coronagraph prioritizes ease of use with well-documented callable functions and intuitive parameter settings. Detailed documentation, automatically generated from Python docstrings, can be accessed online at GitHub repository \footnote{\url{https://dreamjade.github.io/Toy_Coronagraph/}}. In addition to the function reference, an example Jupyter notebook with step-by-step tutorials is available on GitHub \footnote{\url{https://github.com/dreamjade/Toy_Coronagraph/blob/main/tests/test.ipynb}}. These resources provide users with the necessary information and guidance to effectively utilize the package for their research needs.

\end{itemize}

\section{Applications}
By comparing simulations with and without the inclusion of dust inside the \gls{IWA}, Toy Coronagraph highlights the significant underestimation of background noise that can occur when inner dust is neglected. This underscores the importance of understanding the density and structure of exozodiacal dust for accurate predictions of future exoplanet imaging missions.

Toy Coronagraph has broad applications in exoplanet research:
\begin{itemize}
\item \textbf{Understanding Exozodiacal Dust Impact:} The tool enables in-depth studies of how dust affects observations, paving the way for informed mitigation strategies.
\item \textbf{Developing Mitigation Methods:} By simulating various scenarios, researchers can develop and test techniques to minimize the impact of dust on exoplanet detection.
\item \textbf{Future Space Telescope Design Optimization:} Toy Coronagraph can inform the design of future telescopes and coronagraphs to maximize their effectiveness in the presence of exozodiacal dust.
\item \textbf{Mission Time Allocation Optimization:} By accurately predicting the impact of dust, the tool can aid in optimizing observation strategies and maximizing the scientific return of future missions.
\end{itemize}

\section{Code Availability and Usage}

The Toy Coronagraph software package is publicly available and can be accessed through both PyPI \footnote{\url{https://pypi.org/project/toycoronagraph/}} and GitHub \footnote{\url{https://github.com/dreamjade/Toy_Coronagraph}}. Researchers can readily install the package using either of the following methods: 

\begin{itemize}
\item Cloning the whole repository (execute the command below to obtain a local copy of the source code):

\$ git clone \url{https://github.com/dreamjade/Toy_Coronagraph.git}

\item Installation via pip (employ the command below for a streamlined installation process):

\$ pip install toycoronagraph

\end{itemize}

The codebase has been archived and assigned a DOI via Zenodo \cite{yu_chia_lin_2023_8350384} to ensure proper citation and long-term accessibility. The author encourages the scientific community to utilize Toy Coronagraph for investigating the impact of exozodiacal dust on exoplanet detection and characterization efforts. Insights gained from these studies will be crucial for advancing our understanding of habitable worlds beyond our solar system.

\acknowledgments 

The author acknowledges valuable inputs from Ewan S. Douglas, Justin Hom, Ramya M. Anche, and the rest of the AstroCode and UASAL teams. This work was supported by the University of Arizona and used University of Arizona High-Performance Computing (HPC) resources. This research made use of community-developed core Python packages, including Astropy \cite{2022ApJ...935..167A}, ffmpeg-python \cite{ffmpeg}, HCIPy \cite{por2018hcipy}, Matplotlib \cite{Hunter:2007}, Numpy \cite{2020NumPy-Array}, Pysynphot \cite{Lisa_My_Research_Software_2017}, Scikit-image \cite{van2014scikit}, SciPy \cite{scipy}, tqdm \cite{casper_da_costa_luis_2024_11107065}, the IPython Interactive Computing architecture \cite{4160251}, and Jupyter \cite{2016ppap.book...87K}.  I acknowledge the use of GitHub Copilot, an AI-powered code completion tool, to assist in generating Python docstrings. Additionally, Google AI Studio was employed to improve the readability of this proceeding.

\bibliography{report} 
\bibliographystyle{spiebib} 

\end{document}